# Argon assisted chemical vapor deposition of CrO$_2$: an efficient process leading to high quality epitaxial films

**A.C. Duarte[a,b], N. Franco[c], A.S. Viana[d], N.I. Polushkin[a,b], A.J. Silvestre[b,e], O. Conde[a,b*]**

[a]Departamento de Física, Faculdade de Ciências, Universidade de Lisboa, 1749-016 Lisboa, Portugal

[b]Centro de Física e Engenharia de Materiais Avançados (CeFEMA), 1049-001 Lisboa, Portugal

[c]Centro de Ciências e Tecnologias Nucleares, IST, Universidade de Lisboa, 2695-066 Bobadela, Portugal

[d]Centro de Química e Bioquímica, Faculdade de Ciências, Universidade de Lisboa, 1749-016 Lisboa, Portugal

[e]Área Departamental de Física, ISEL – Instituto Superior de Engenharia de Lisboa, Instituto Politécnico de Lisboa, 1959-007 Lisboa, Portugal

## Abstract

A comparative study of the structural, microstructural and magnetic properties of CrO$_2$ thin films grown onto (110) and (100) TiO$_2$ rutile single crystal substrates by chemical vapor deposition (CVD), using CrO$_3$ as chromium precursor and either oxygen or argon as carrier gas is presented. Our results show that growth under argon carrier gas leads to high quality CrO$_2$ epilayers with structural and magnetic properties similar to those obtained using the more standard oxygen carrier gas. Furthermore, we interpret the larger magnetic coercivity observed for the (110) oriented films in terms of their microstructure, in particular of the highest strain and edge roughness of the building structures of the CrO$_2$ epilayers, which are settled by the substrate crystallographic orientation.

---

*Corresponding author: e-mail: omconde@ciencias.ulisboa.pt

# 1. Introduction

$CrO_2$ is a material of choice for spintronics. It has been extensively studied by the scientific community [1-12] since the pioneering work carried out in the 60's of last century as reviewed in ref. [13]. Recent findings such as the spin triplet supercurrents [14-17], the coexistence of universal and topological anomalous Hall effects and the robustness of spin polarization up to room temperature [18] have strongly contributed to keep alive, or even increase, the interest on potentially using $CrO_2$ thin films in spintronic and magnonic devices [19-22]. But the studies on chromium dioxide have not been limited to fundamental or applied science. Indeed, $CrO_2$ has been used in industry covering a wide range of applications *e.g.* audio, video, instrumentation and computer technology due to its exceptional properties as a magnetic storage material [13].

Although some attempts have been pursued to grow thin films of the single $CrO_2$ compound by physical vapor deposition methods [23-25], only chemical vapor deposition (CVD) has shown to lead to high quality $CrO_2$ layers [26-30]. $CrO_2$ epitaxial layers are currently grown on $TiO_2$ rutile phase and sapphire substrates by CVD using $CrO_3$ as the chromium precursor, which is carried out into the reaction zone by an oxygen flux. While chromium precursors other than $CrO_3$, *e.g.* $Cr_8O_{21}$ [31] and $CrO_2Cl_2$ [8,10], have also been used in the CVD of $CrO_2$ yielding films with similar properties, the use of an oxygen flux has been pointed out as critical for the overall $CrO_2$ synthesis process. To our knowledge, only the work of Ivanov *et al.* [31] briefly refers that the use of argon as carrier gas makes the process very inefficient and leads to films with poorer crystallinity.

Currently, the demand for fast and cost-efficient fabrication processes of high-quality materials and devices is on the agenda. Hence, an in-depth study of the feasibility of growing high quality epitaxial films of $CrO_2$ using argon as carrier gas becomes crucial, especially because of the well-known price difference between the two gases, with cost advantage to argon. Besides, argon is a more user-friendly gas. Therefore, the fact that the use of argon could eventually limit the $CrO_2$ synthesis reaction may be seen as a drawback to the envisaged scaling-up of the $CrO_2$ CVD process at an industrial level.

This work reports on the synthesis and properties of $CrO_2$ thin films grown onto single crystal $TiO_2$ rutile substrates by CVD, using $CrO_3$ as chromium precursor and either oxygen or argon as carrier gas. A comparative study of the structural, microstructural and magnetic properties of the different films are presented.

## 2. Experimental details

The $CrO_2$ films were prepared in a tubular single-zone CVD furnace with independent control of the substrate temperature. Details of this setup were described in previous papers [28,29]. The thin films were grown onto $TiO_2$ (110) and $TiO_2$ (100) single crystal rutile substrates 0.5 mm thick and with $10\times5$ mm$^2$ surface area. Prior to their insertion into the reactor, the substrates were ultrasonically cleaned in organic solvents – acetone and isopropanol, rinsed in distilled water and dried with a $N_2$ flux. $CrO_3$ powder (purity 99.9%) was used as the precursor and either oxygen (purity 99.9992%) or argon (purity 99.9992%) as carrier gases. Films were grown at atmospheric pressure for deposition times varying between 10 and 240 min, using a gas flow rate of 150 sccm, a precursor temperature of 275 ºC and a substrate temperature of 390 ºC. In order to avoid the deposition of undesirable compounds during the initial stage of the deposition process, the substrate was heated up to the deposition temperature before the $CrO_3$ precursor attained its melting temperature (196 ºC). Samples will be designated by A100, A110, O100 and O110 throughout the text, where A/O refers to argon/oxygen and 100/110 to the substrate orientation. Sometimes the deposition time, in minutes, is added to the sample's label (e.g. O110/180).

The surface morphology and microstructure of the films were analyzed using a JEOL 7001F scanning electron microscope with a field emission gun (FEG-SEM), and a Nanoscope

IIIa multimode atomic force microscope (AFM) in tapping mode. Film thicknesses were evaluated by image digital processing of FEG-SEM cross-section micrographs. The crystallographic structure and phase composition of the as-grown films were studied by X-ray diffraction (XRD) with Cu K$\alpha$ radiation, in the $\theta$–$2\theta$ coupled mode. The $2\theta$ angular position and full-width at half-maximum (FWHM) of the diffraction lines were determined by fitting them with a Pseudo-Voigt function, after removing the $K\alpha_2$ contribution from the XRD patterns. The FWHM values were also corrected for the instrumental broadening. Phi-scan and rocking curve (RC) measurements were used to assess films epitaxiality and mosaicity. These studies were conducted on a Brücker-AXS D5000 diffractometer. Another diffractometer from Brücker-AXS (D8Discover) working with Cu K$\alpha_1$ radiation, an asymmetric two-bounce Ge (220) monochromator and a scintillation detector was used for recording the reciprocal space maps (RSM) and for carrying out the evaluation of the lattice parameters using the Bond method.

The magnetic properties of all the films were assessed at room temperature (RT) with longitudinal magneto-optical Kerr effect (MOKE) magnetometry. To observe the MOKE, the probing beam from a diode laser, operating at a wavelength of 670 nm with a power of 5.5 mW, was focused onto a ~50 μm diameter spot on the film surface using a 15 cm focal length lens.

## 3. Results and discussion

### 3.1 Structural data

The first set of results concerns XRD analysis of chromium oxide films grown on $TiO_2$ (110) substrates, using either oxygen or argon as carrier gas. This study is aimed at assessing the feasibility of $CrO_2$ growth under an argon atmosphere and, if the latter holds, at comparing film properties for both types of background gas phase. Of particular importance is to find out whether the presence of argon does or does not favour the occurrence of the antiferromagnetic $Cr_2O_3$ phase.

Fig. 1 shows the XRD patterns of films deposited at 390 °C with an oxygen flow rate of 150 sccm, for deposition times in the range of 60 to 240 minutes. All the patterns reveal diffraction peaks at $2\theta = 28.643°$, $2\theta = 59.096°$ and $2\theta = 95.335°$, which were identified as due to the (110), (220) and (330) reflections from $CrO_2$, respectively, by comparing with JCPDS file nº 9-332. The diffraction peaks labelled with the symbol (*) are attributed to the (110), (220) and (330) reflections from the rutile-$TiO_2$ substrate. No other phases are visible on this figure even if a logarithmic scale is used to plot the diffracted intensities. Similar patterns are displayed in Fig. 2a for the films deposited with argon substituting for oxygen as carrier gas. The growth parameters are identical for the two gases as well as the deposition times that were used. The same set of diffraction peaks related to the (110), (220) and (330) lattice planes of both $CrO_2$ film and rutile-$TiO_2$ substrate are displayed, however a low intensity extra peak may be observed on the low angle shoulder of the $TiO_2$ (220) line that can be assigned to the $Cr_2O_3$ (116) diffraction plane (JCPDS file nº 38-1479). This peak is clearly visible on an expanded $\theta-2\theta$ plot as that in Fig. 2b. It is present in all films as a very broad and low intensity peak indicating that the $Cr_2O_3$ formed phase is nanocrystalline and its amount is negligible. Nevertheless, the increase in intensity as the deposition time decreases, i.e. as the films become thinner, might be an indication that the $Cr_2O_3$ lies at the interface between the $CrO_2$ layer and the substrate.

The X-ray diffractograms of the films grown on $TiO_2$ (100) for a deposition time of 90 minutes are displayed in Fig. 3, for both types of carrier gas – oxygen and argon. Here, all the experimental parameters were maintained constant and assumed the same values as for the previous (110) oriented substrate. The $CrO_2$ layers grow with a (100) preferred orientation according to the orientation of the substrate and diffraction lines attributed to the (200) and

(400) reflections of CrO$_2$, respectively at $2\theta = 40.855°$ and $2\theta = 88.491°$, and of TiO$_2$ were recorded. Besides these diffraction peaks, two other ones at $2\theta = 36.196°$ and $2\theta = 76.851°$ are clearly observed, which were indexed as (110) and (220) reflections from the Cr$_2$O$_3$ phase. This indicates that the Cr$_2$O$_3$ layer is textured and strongly correlated with the CrO$_2$ (100) – TiO$_2$ (100) system.

Indeed, it is well known that CrO$_2$ can grow epitaxially on TiO$_2$ substrates because both compounds adopt the same tetragonal rutile structure ($P4_2/mnm$) with a close match in lattice parameters: $a=b=0.45933$ nm and $c=0.29592$ nm for TiO$_2$, and $a=b=0.44210$ nm and $c=0.29160$ nm for CrO$_2$. Therefore, the lattice mismatch is $-1.46\%$ along the [001] direction ($c$ axis) and $-3.75\%$ along either the [010] axis for TiO$_2$ (100) or the [$\bar{1}$10] direction for TiO$_2$ (110) substrate. In both cases CrO$_2$ films grow under interfacial tensile stress. Regarding the presence of the Cr$_2$O$_3$ layer in the films with (100) preferred orientation, it is expected that this layer will grow on top of CrO$_2$ since the surface area of the cells match to 99.4%, although 3×3 cells of CrO$_2$ are needed to fit one cell of Cr$_2$O$_3$, whereas the corresponding matching value for Cr$_2$O$_3$ directly grown on 3×3 TiO$_2$ (100) would be only 95.4%. It is believed that the Cr$_2$O$_3$ layer forms during the cooling of the samples inside the CVD reactor, from the deposition temperature of 390 °C down to room temperature. Using the method described in ref. [29], the thickness of this Cr$_2$O$_3$ top layer can be calculated. For instance, a value of 24.8 nm was obtained for the film A100/90, which represents 11.8% of the total film thickness.

In order to assess the crystalline quality of the films and possible structural misalignments between film and substrate, reciprocal space maps (RSM) were recorded around the (220) and (200) reflections of CrO$_2$ for the samples grown on TiO$_2$ (110) and TiO$_2$ (100), respectively. Figs. 4a-4d displays typical symmetric RSMs for films of the different types although keeping the deposition time constant at 90 minutes. A qualitative inspection of these maps shows a strong coherence between the [110] axes of $CrO_2$ and $TiO_2$ in case of the (110) oriented substrate (Figs. 4a and 4c). For the (100) $CrO_2/TiO_2$ (Figs. 4b and 4d) a strong coherence is also observed not only between the [100] axis of both materials as well as the [110] axis of $Cr_2O_3$, which reinforces our previous statement on the structural relationship of the three compounds. Therefore, the following in-plane orientation epitaxial relationships can be written: $CrO_2(110)[110] \mid TiO_2(110)[110]$ and $Cr_2O_3(110)[110] \mid CrO_2(100)[100] \mid TiO_2(100)[100]$.

An easy, rapid and accurate way of comparing these maps amongst themselves will consist of making vertical and horizontal cuts and plotting the intensities versus Qz (Figs. 4e and 4f) and $Q_x$ (Figs. 4g and 4h), respectively. The first plot gives the lattice parameter $a$ for the films and substrates, determined from $a = 2\pi(h^2 + k^2)^{1/2}/Q_z$. The second plot is related to the crystalline quality of the $CrO_2$ layers, assessed through the analysis of the FWHM of the profiles. In general, the lines are broader (along $Q_x$) and thus a more pronounced mosaic structure is observed for the films grown under oxygen. This result agrees well with the rocking curves (not shown) measured around the (110) reflection, whose FWHM decreases from 1.2º down to 0.8º when deposition time increases from 60 to 240 minutes, the values for oxygen exceeding in general the argon ones. More striking is the difference of RC for both film orientations, for instance at $t_{dep}$=90 min, we measured 1.0º and 0.5º for (110) and (100) films, respectively. Table 1 summarizes the $Q_z$ and lattice parameter $a$ values for different samples. While the reproducibility is very high regarding the $TiO_2$ substrates, it is apparent that the $CrO_2$ films grown with (110) preferred orientation are under tensile stress (tensile strain <0.3%), assuming that a relaxed film will show $a = 0.4421$ nm. A calculated error of $2 \times 10^{-4}$ nm is associated to the lattice parameter $a$ of $CrO_2$.

Neither the $\theta$–$2\theta$ coupled diffraction nor the symmetric RSMs allow determining the $c$ parameter. The lattice parameter $c$ was determined with a high accuracy by using an XRD technique based on the Bond method [32,33] and the asymmetric reflections (112) and (202) respectively for (110) and (100) oriented substrates. It was assumed that $a = b$. We show on Fig. 5 an example of these measurements, and on Table 2 a summary of the results for selected samples. The calculated error associated to the lattice parameter $c$ of $CrO_2$ is $3 \times 10^{-5}$ nm. It comes out from these results that the structural parameters of $CrO_2$ films grown on $TiO_2$ substrates are not sensitive to the carrier gas used for film growth whether it is oxygen or argon. This is true for the (tensile or compressive) stress acting on the films as a result of lattice mismatch and/or difference in the thermal expansion coefficients of film and substrate. On the other hand, the orientation of the substrate strongly influences the phase composition of the deposited material as well as the structural behaviour of the $CrO_2$ films. Indeed, it is clearly seen that while the films with (110) preferred orientation are compressively strained, those with (100) are almost fully relaxed, which allows to explain why their FWHM-RC are the lowest. This notorious influence of the substrate is in agreement with previous observations from other research groups however, while Anwar and Aarts [34] have found a similar behaviour to our films, Chetry *et al.* [35] and Pathak *et al.* [36] found the opposite and based their interpretation on differences in growth mode and morphology of the $CrO_2$ films. We will come back to this point in section 3.3.

**3.2 Film thickness and growth rate**

The $CrO_2$ growth kinetics depends on a number of factors such as the geometry of the reactor, substrate material and crystallographic orientation, substrate and precursor temperatures, carrier gas, and carrier gas flow rate [28,29]. This section will focus on the role of the carrier gas on the thickness and growth rate of $CrO_2$ films for both types of single crystal substrates. Fig. 6 shows the thickness of $CrO_2$ films grown onto $TiO_2$ (110) as a function of deposition time for both oxygen and argon carrier gases. The inset of the figure shows the thickness of the films grown onto $TiO_2$ (100) for deposition times of 90 and 120 minutes. First of all, we notice that films grown with an oxygen flux are slightly thicker than those grown with an argon flux regardless of substrate orientation. Only at $t_{dep} = 240$ min is this trend reversed, which is probably related to the faster consumption of the $CrO_3$ precursor when using oxygen carrier gas. Secondly, the thickness of films grown onto $TiO_2$ (100) is larger than that of films grown onto $TiO_2$ (110) by a factor of about 1.7−1.8. This value is slightly larger than the interplanar spacing ratio, $d_{100}/d_{110} = \sqrt{2}$, between (100) and (110) lattice planes of a tetragonal structure. Thus, the preferred growth orientation is in part responsible for the difference in thickness observed for the films grown on the two different substrates. Besides, the formation of a $Cr_2O_3$ layer on top of $CrO_2$ (100) films and different growth morphology (see below) can also be responsible for the observed thickness difference.

The thickness of $CrO_2(110)$ films *vs.* deposition time (Fig. 6) shows two different regions − a linear region that extends up to ~100 minutes and a saturation region for deposition times greater than ~140 minutes. In the linear region, growth rates of $1.10 \pm 0.04$ nm and $1.10 \pm 0.07$ nm can be deduced for the films grown respectively with $O_2$ and Ar carrier gases. This result clearly shows that film growth is not influenced by which gas is used to transport the Cr precursor to the substrate. The observed transition from the linear stage to the saturation regime is due to a decrease of the Cr precursor concentration in the gas phase.

### 3.3 Morphology

The surface microstructure and morphology of the as-grown films was examined by FEG-SEM and AFM and is depicted in Fig. 7 for 90 minutes deposited films. Again, the type of background gas does not have a perceptible effect on film morphology and only the substrate orientation affects the film growth mode. For both types of substrates (and gases) similar $CrO_2$ building blocks consisting of rectangular shaped grains were observed. However, the way these grains pile up to grow a film varies with substrate orientation. For CrO$_2$ (100) the stacking of rectangular plates on top of each other with the same orientation is clearly observed on the AFM insets. This microstructural arrangement leaves low-density boundaries between the building blocks, or even voids, which allow explaining the smaller values of the FWHM-rocking curves, the almost absence of strain and rather smooth edges of the structures seen on the SEM images. On the contrary, for the CrO$_2$ (110) samples the piling-up of the rectangular platelets is not so perfect as previously and some misorientation exists, which leads to higher FWHM-RC values and denser structures with increased edges roughness. The root mean square surface roughness ($R_q$) measured on the insets (2×2 μm$^2$) of Fig. 7 took values within 2.8 – 3.9 nm and 5.4 – 6.5 nm respectively for CrO$_2$ (110) and CrO$_2$ (100), in agreement with the observed microstructures.

### 3.4 Magnetic measurements - Coercivity

We have used longitudinal MOKE magnetometry with the applied magnetic field parallel to the easy axis (*c*-axis, oriented along the larger side of the rectangular grains) to measure the room temperature magnetic properties of the as-deposited CrO$_2$ films on (110) and (100) TiO$_2$ substrates with oxygen and argon as carrier gases. Fig. 8 shows the hysteresis loops of samples deposited for 90 minutes, where it is observed an increase of ≤ 15% in maximum MOKE response when (110) substrate is replaced by the (100) one, under the same background gas. A similar increase is also observed for both substrates when oxygen is replaced by argon as flowing gas. On the contrary, a sharp reduction of ~ 60% in the switching field is observed for the (100) films grown during 90 minutes in comparison to the (110) ones. However, if the comparison is made between films of different orientation but the same thickness then the switching field reduces by ~ 40%. The carrier gas seems to have no remarkable influence on the coercive field. Various factors can contribute to the change of coercivity, such as stoichiometry deviation, grain size and strain [37]. Because Scherrer equation [38] applied to the analysis of the XRD peaks gives similar grain size for the films grown with either oxygen or argon, the highest strain observed for the (110) films should be the main cause for their enhanced coercivity. On the other hand, structural imperfections, such as edge roughness, have been shown to alter the switching field of elongated structures [39], the coercivity increasing with the increase in edge roughness. For our films, the SEM and AFM images on Fig. **7** clearly show rough edges of the (110) elongated structures whereas the (100) blocks present sharp and smoother edges as discussed above in section 3.3.

**Conclusion**

CrO$_2$ thin films were grown onto (110) and (100) TiO$_2$ rutile single crystal substrates by CVD using a highly efficient tubular single-zone CVD furnace with independent control of the substrate temperature. CrO$_3$ was used as chromium precursor and either O$_2$ or Ar as carrier gas. It is shown that the use of an Ar flux rather than an O$_2$ flux is not a limitation regarding the deposition of CrO$_2$. Instead, the use of argon carrier gas allows to grow high quality CrO$_2$ epilayers with similar structural, microstructural and magnetic properties of those grown with O$_2$ carrier gas. Also, the growth rate of the films does not depend on the carrier gas used. These results contrast with those reported by Ivanov et al., who mentioned [31] a low efficient growth process and poor quality films for argon assisted CVD CrO$_2$. For each type of carrier gas, the properties of the films are mainly determined by substrate orientation.

**Acknowledgments**

This work was supported by the Portuguese Foundation for Science and Technology (FCT), through research contracts PTDC/FIS/121588/2010, PEst-OE/CTM/UI-00084 and UID/CTM/04540/2013. N.I.P. and A.S.V. acknowledge financial support from FCT within program "Ciência 2008" and "IF2013 Initiative (POPH, FSE)", respectively.

# References

[1] H.Y. Hwang, S.-W. Cheong, Enhanced intergrain tunneling magnetoresistance in half-metallic $CrO_2$ films, Science 278 (1997)1607−1609.

[2] L. Ranno, A. Barry, J.M.D. Coey, Production and magnetotransport properties of $CrO_2$ films, J. Appl. Phys. 81 (1997) 5774−5776.

[3] J.M.D. Coey, A.E. Berkowitz, L. Balcells, F.F. Putris, Magnetoresistance of chromium dioxide powder compacts, Phys. Rev. Lett. 80 (1998) 3815−3818.

[4] K. Suzuki, P.M. Tedrow, Resistivity and magnetotransport in $CrO_2$ films, Phys. Rev. B 58 (1998) 11597−11602.

[5] X.W. Li, A. Gupta, T.R. McGuire, P.R. Duncombe, G. Xiao, Magnetoresistance and Hall effect of chromium dioxide epitaxial thin films, J. Appl. Phys. 85 (1999) 5585−5587.

[6] A. Barry, J.M.D. Coey, M.A. Viret, $CrO_2$-based magnetic tunnel junction, J. Phys.: Condens. Matter 12 (2000) L173−L175.

[7] M. Rabe, J. Dreβen, D. Dahmen, J. Pommer, H. Stahl, U. Rüdiger, G. Güntherodt, S. Senz, D. Hesse, Preparation and characterization of thin ferromagnetic $CrO_2$ films for applications in magnetoelectronics, J. Magn. Magn. Mater. 211 (2000) 314−319.

[8] W.J. DeSisto, P.R. Broussard, T.F. Ambrose, B.E. Nadgorny, M.S. Osofsky, Highly spin-polarized chromium dioxide thin films prepared by chemical vapor deposition from chromyl chloride, Appl. Phys. Lett. 76 (2000) 3789−3791.

[9] P.A. Stampe, R.J. Kennedy, S.M. Watts, S.v. Molnár, Strain effects in thin films of $CrO_2$ on rutile and sapphire substrates, J. Appl. Phys. 89 (2001) 7696−7698.

[10] A. Anguelouch, A. Gupta, G. Xiao, D.W. Abraham, Y. Ji, S. Ingvarsson, C.L. Chien, Near-complete spin polarization in atomically-smooth chromium-dioxide epitaxial films prepared using a CVD liquid precursor, Phys. Rev. B 64 (2001) 180408(R).

[11] U. Rüdiger, M. Rabe, K. Samm, B. Özyilmaz, J. Pommer, M. Fraune, G. Güntherodt, S. Senz, D. Hesse, Extrinsic and intrinsic magnetoresistance contribuitions of $CrO_2$ thin films, J. Appl. Phys. 89 (2001) 7699−7701.

[12] Y.S. Dedkov, M. Fonine, C. König, U. Rüdiger, G. Güntherodt, Room-temperature observation of high-spin polarization of epitaxial $CrO_2$ (100) island films at the Fermi energy, Appl. Phys. Lett. 80 (2002) 4181−4183.

[13] V.A. Jaleel, T.S. Kannan, Hydrothermal synthesis of chromium dioxide powders and their characterization, Bull. Mater. Sci. 5 (1983) 231−246.

[14] R.S. Keizer, S.T.B. Goennenwein, T.M. Klapwijk, G. Miao, G. Xiao, A. Gupta, A spin triplet supercurrent through the half-metallic ferromagnet $CrO_2$, Nature 439 (2006) 825−827.

[15] M.S. Anwar, M. Veldhorst, A. Brinkman, J. Aarts, Long range supercurrents in ferromagnetic $CrO_2$ using a multilayer contact structure, Appl. Phys. Lett. 100 (2012) 052602.

[16] K.A. Yates, W.R. Branford, F. Magnus, Y. Miyoshi, B. Morris, L.F. Cohen, P.M. Sousa, O. Conde, A.J. Silvestre The spin polarization of $CrO_2$ revisited, Appl. Phys. Lett. 91 (2007) 172504.

[17] A. Singh, S. Voltan, K. Lahabi, J. Aarts, Colossal proximity effect in a superconducting triplet spin valve based on the half-metallic ferromagnet $CrO_2$, Phys. Rev. X 5 (2015) 021019.

[18] W.R. Branford, K.A. Yates, E. Barkhoudarov, J.D. Moore, K. Morrison, F. Magnus, Y. Miyoshi, P.M. Sousa, O. Conde, A.J. Silvestre, L.F. Cohen, Coexistence of universal and topological anomalous hall effects in metal $CrO_2$ thin films in the dirty limit, Phys. Rev. Lett. 102 (2009) 227201.

[19] Y. Ding, C. Yuan, Z. Wang, S. Liu, J. Shi, R. Xiong, D. Yin, Z. Lu, Improving thermostability of $CrO_2$ thin films by doping with Sn, Appl. Phys. Let. 105 (2014) 092401.

[20] K.G. West, M. Osofsky, I.I. Mazin, N.N.N. Dao, S.A. Wolf, J. Lu, Magnetic properties and spin polarization of Ru doped half metallic $CrO_2$, Appl. Phys. Let. 107 (2015) 12402.

[21] I.V. Solovyev, I.V. Kashin, V.V. Mazurenko, Mechanisms and origins of half metallic ferromagnetism in $CrO_2$, Phys. Rev. B 92 (2015) 144407.

[22] S. Mironov, A. Buzdin, Triplet proximity effect in superconducting heterostructures with a half metallic layer, Phys. Rev. B 92 (2015) 184506.

[23] M. Shima, T. Tepper, C.A. Ross, Magnetic properties of chromium oxide and iron oxide films produced by pulsed laser deposition, J. Appl. Phys. 91 (2002) 7920−7922.

[24] N. Popovici, M.L. Paramês, R.C. Silva, O. Monnereau, P. Sousa, A.J. Silvestre, O. Conde, KrF pulsed laser deposition of chromium oxide thin films from $Cr_8O_{21}$ targets,

Appl. Phys. A 79 (2004) 1409–1411.

[25] D. Stanoi, G. Socol, C. Grigorescu, F. Guinneton, O. Monnereau, L. Tortet, T. Zhang, I.N. Mihailescu, Chromium oxides thin films prepared and coated in situ with gold by pulsed laser deposition, Mater. Sci. Eng. B 118 (2005) 74–78.

[26] X.W. Li, A. Gupta, G. Xiao, Influence of strain on the magnetic properties of epitaxial (100) chromium dioxide ($CrO_2$) films, Appl. Phys. Lett. 75 (1999) 713–715.

[27] M. Rabe, J. Pommer, K. Samm, B. Özyilmaz, C. König, M. Fraune, U. Rüdiger, G. Güntherodt, S. Senz, D. Hess, Growth and magnetotransport study of thin ferromagnetic $CrO_2$ films, J. Phys.: Condens. Matter 14 (2002) 7–20.

[28] P.M. Sousa, S.A. Dias, A.J. Silvestre, O. Conde, B. Morris, K.A. Yates, W.R. Branford, L.F. Cohen, CVD of $CrO_2$: Towards a lower temperature deposition process, Chem. Vapor Deposition 12 (2006) 712–714.

[29] P.M. Sousa, S.A. Dias, O. Conde, A.J. Silvestre, W.R. Branford, B. Morris, K.A. Yates, L.F. Cohen Influence of growth temperature and carrier flux on the structure and transport properties of highly oriented $CrO_2$ on $Al_2O_3$ (0001), Chem. Vapor Deposition 13 (2007) 537–545.

[30] C. Aguilera, J.C. González, A. Borrás, D. Margineda, J.M. González, A.R. González-Elipe, J.P. Espinós, Preparation and characterization of $CrO_2$ films by Low Pressure Chemical Vapor Deposition from $CrO_3$, Thin Solid Films 539 (2013) 1–11.

[31] P.G. Ivanov, S.M. Watts, D.M. Lind, Epitaxial growth of $CrO_2$ thin films by chemical-vapor deposition from a $Cr_8O_{21}$ precursor, J. Appl. Phys. 89 (2001) 1035–1040.

[32] W.L. Bond, Precision lattice constant determination, Acta Cryst. 13 (1960) 814–818.

[33] N. Herres, L. Kirste, H. Obloh, K. Koehler, J. Wagner, P. Koidl, X-ray determination of the composition of partially strained group-III nitride layers using the extended Bond method, Mater. Sci. and Eng. B 91–92 (2002) 425–432.

[34] M.S. Anwar, J. Aarts, Inducing supercurrents in thin films of ferromagnetic $CrO_2$, Supercond. Sci. Technol. 24 (2011) 024016.

[35] K.B. Chetry, M. Phatak, P. LeClair, A. Gupta, Structural and magnetic properties of (100)- and (110)-oriented epitaxial $CrO_2$ films. J. Appl. Phys. 105 (2009) 083925.

[36] M. Phatak, H. Sato, X. Zhang, K.B. Chetry, D. Mazumdar, P. LeClair, A. Gupta, Substrate-induced strain and its effect in $CrO_2$ thin films, J. Appl. Phys. 108 (2010)

053713.

[37] N. Li, A.Y.-H.A. Wang, N.M. Iliev, T.M. Klein, A. Gupta, Growth of atomically smooth epitaxial nickel ferrite films by direct liquid injection CVD, Chem. Vap. Deposition 17 (2011) 261–269.

[38] B.D. Cullity, S.R. Stock, Elements of x-ray diffraction, 3rd edition, Prentice Hall, New Jersey, 2001, p. 170.

[39] M.T. Bryan, D. Atkinson, R.P. Cowburn, Edge roughness and coercivity in magnetic nanostructures. J. Phys.: Conf. Series 17 (2005) 40–44, and references therein.

**Table Captions**

**Table 1** $Q_z$ cuts on the symmetrical RSM for the substrate and film, and determination of the lattice parameter $a$ from the following equation: $a = 2\pi(h^2+k^2)^{1/2}/Q_z$.

**Table 2** Lattice parameters $a$ and $c$ of $CrO_2$ thin films, determined from symmetrical RSMs and Bond method, and tensile ($\varepsilon > 0$) / compressive ($\varepsilon < 0$) strain along the [100] and [001] crystallographic directions.

**Figure Captions**

**Fig. 1** X-ray diffractograms of O110-films grown at 390 ℃ with an oxygen flow rate of 150 sccm, for deposition times ranging from 60 to 240 minutes. The patterns show the (hh0) lines of $CrO_2$ while those labeled with * refer to the rutile-$TiO_2$ (110) single crystal substrate.

**Fig. 2** a) X-ray diffractograms of A110-films grown at 390 ℃ with an argon flow rate of 150 sccm for deposition times ranging from 60 to 240 minutes. The patterns show the (hh0) lines of $CrO_2$ while those labeled with * refer to the rutile-$TiO_2$ (110) single crystal substrate; b) expanded plot in the *2θ*: 54º – 60º region showing a broad and weak peak assignable to the (116) reflection from $Cr_2O_3$.

**Fig. 3** XRD patterns of films deposited on $TiO_2$ (100) substrate using either argon or oxygen as carrier gas, for a deposition time of 90 minutes. The symbol (*) corresponds to substrate peaks. Both films were deposited with the gas flowing at 150 sccm and for a deposition temperature of 390ºC.

**Fig. 4** Symmetric reciprocal space maps of films deposited with oxygen (panels a) and b)) and argon (panels c) and d)) onto rutile-$TiO_2$ substrates with (110) orientation (top) and (100) orientation (bottom). For all cases, deposition time is 90 minutes. e), f): Vertical ($Q_z$) cut and g), h): Horizontal ($Q_x$) cut taken over the SRSMs shown in a) - d).

**Fig. 5** Schematic configuration (top panel) for the measurement of the XRD asymmetric reflections by the Bond method, which are shown in the lower panel for two films grown with argon on (110) and (100) $TiO_2$ substrates.

**Fig. 6** Thickness *vs.* deposition time for $CrO_2$ films grown on $TiO_2$ (110) substrates with oxygen (full squares) and argon (full circles) carrier gases. The lines drawn are linear best fits. Inset: similar data for $TiO_2$ (100) substrate and deposition times of 90 and 120 minutes.

**Fig. 7** FEG-SEM images of A100/90, O100/90, A110/90 and O110/90 $CrO_2$ films. Insets show the corresponding AFM images with 2×2 $\mu m^2$ and the same *Z* scale (0 – 20 nm).

**Fig. 8** Hysteresis loops of $CrO_2$ films obtained by longitudinal MOKE magnetometry with the applied magnetic field parallel to the easy axis. Substrate orientation, deposition time, and carrier gas are indicated in the figure.

Table **1**

| Sample | TiO$_2$ Q$_z$ (nm$^{-1}$) | CrO$_2$ Q$_z$ (nm$^{-1}$) | TiO$_2$ $a$ (nm) | CrO$_2$ $a$ (nm) |
|---|---|---|---|---|
| O110/180 | 38.699 | 40.102 | 0.45923 | 0.4432 |
| A110/180 | 38.696 | 40.102 | 0.45926 | 0.4432 |
| O110/90 | 38.697 | 40.091 | 0.45925 | 0.4433 |
| A110/90 | 38.697 | 40.091 | 0.45925 | 0.4433 |
| O100/90 | 27.360 | 28.440 | 0.45929 | 0.4419 |
| A100/90 | 27.364 | 28.440 | 0.45923 | 0.4419 |

Table **2**

| Sample | $\omega^+$ | $\omega^-$ | $\theta_B$ [a] | $a$ (nm) [b] | $c$ (nm) | $\varepsilon_{[100]} = \delta a/a$ [c] | $\varepsilon_{[001]} = \delta c/c$ [c] |
|---|---|---|---|---|---|---|---|
| O110/180 (112) | 35.6662 | 143.9174 | 35.8744 | 0.4432 | 0.28094 | $2.49 \times 10^{-3}$ | $-3.65 \times 10^{-2}$ |
| A110/180 (112) | 35.6944 | 143.9389 | 35.8778 | 0.4432 | 0.28091 | $2.49 \times 10^{-3}$ | $-3.67 \times 10^{-2}$ |
| O110/90 (112) | 35.6560 | 143.9121 | 35.8674 | 0.4433 | 0.28099 | $2.71 \times 10^{-3}$ | $-3.64 \times 10^{-2}$ |
| A110/90 (112) | 35.6247 | 143.8528 | 35.8859 | 0.4433 | 0.28085 | $2.71 \times 10^{-3}$ | $-3.69 \times 10^{-2}$ |
| O100/90 (202) | 39.2620 | 140.3580 | 39.4520 | 0.4419 | 0.28998 | $-0.45 \times 10^{-3}$ | $-0.55 \times 10^{-2}$ |
| A100/90 (202) | 39.2445 | 140.3550 | 39.4448 | 0.4419 | 0.29000 | $-0.45 \times 10^{-3}$ | $-0.55 \times 10^{-2}$ |

[a] $\theta_B = [\omega^+ + (180 - \omega^-)]/2$

[b] Value calculated in Table 1

[c] $\varepsilon = \delta x/x = (x_{film} - x_{bulk})/x_{bulk}$

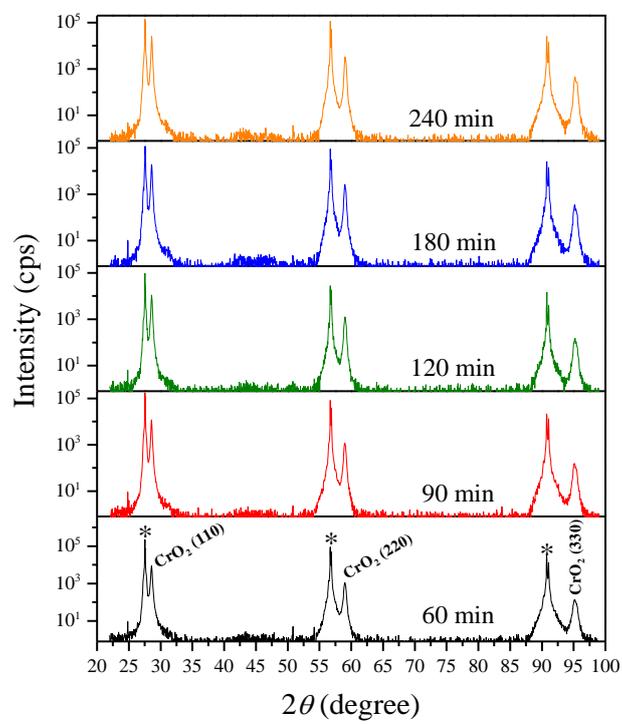

Figure 1

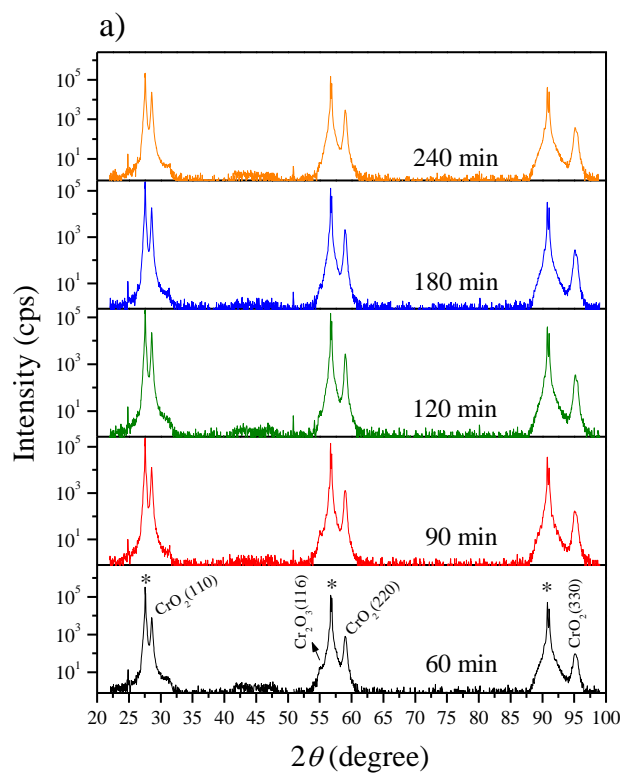

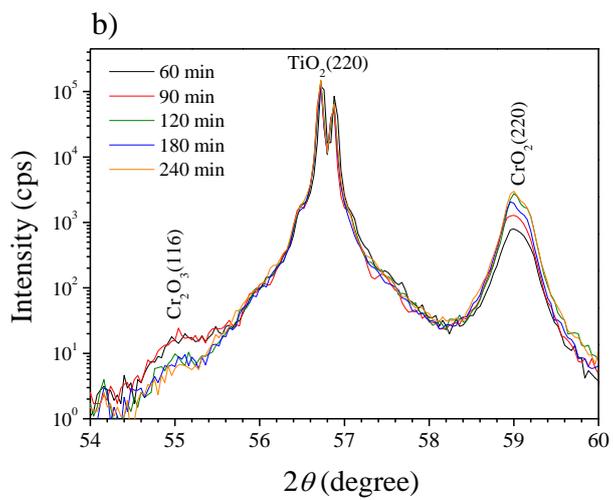

Figure 2

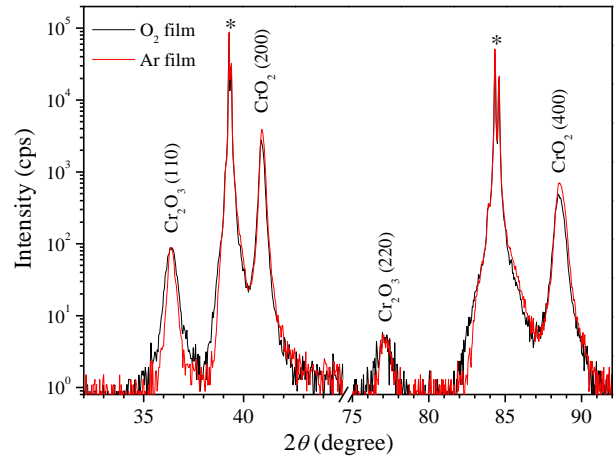

Figure 3

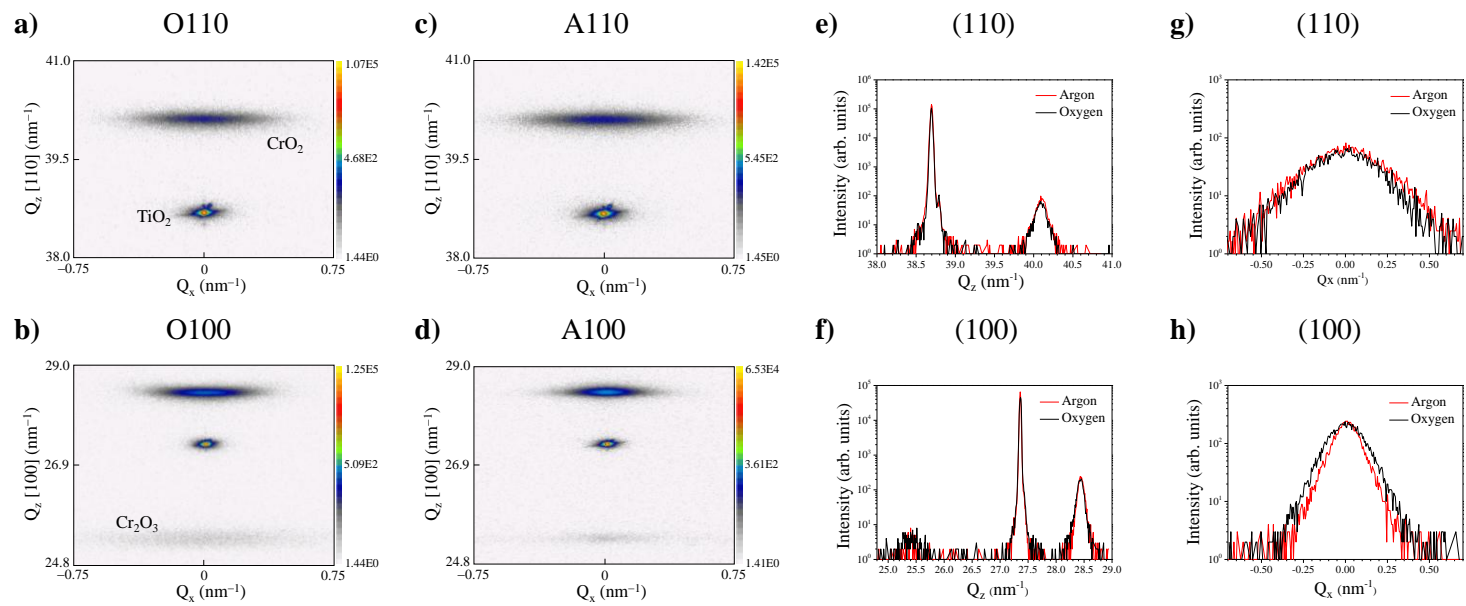

Figure 4

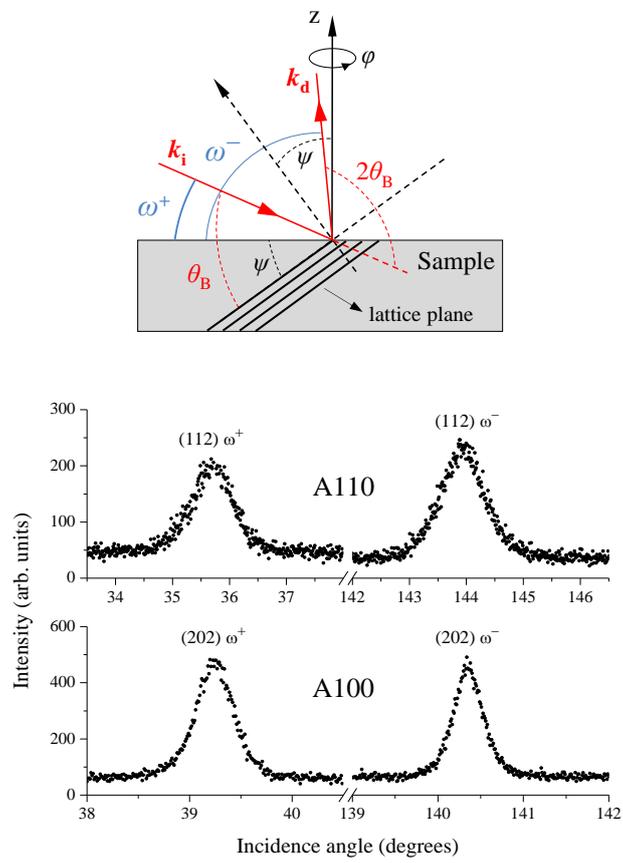

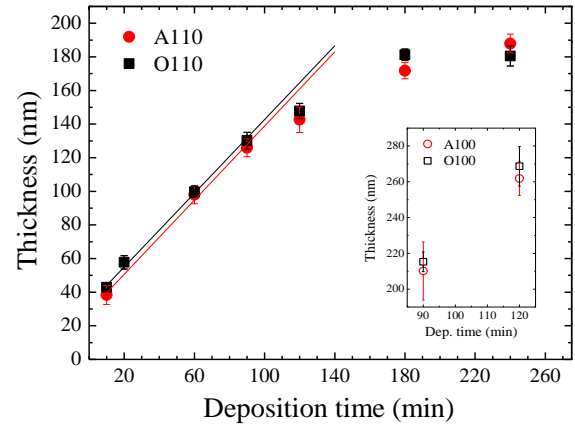

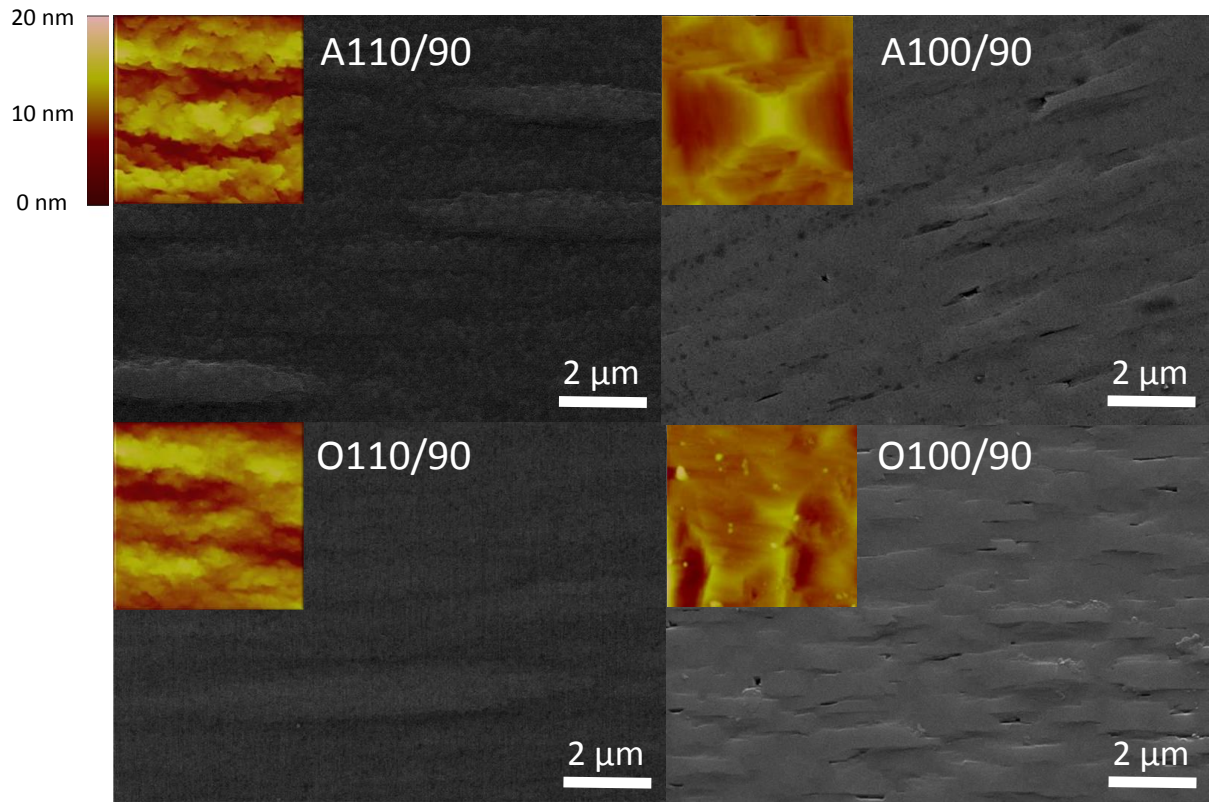

Figure 7

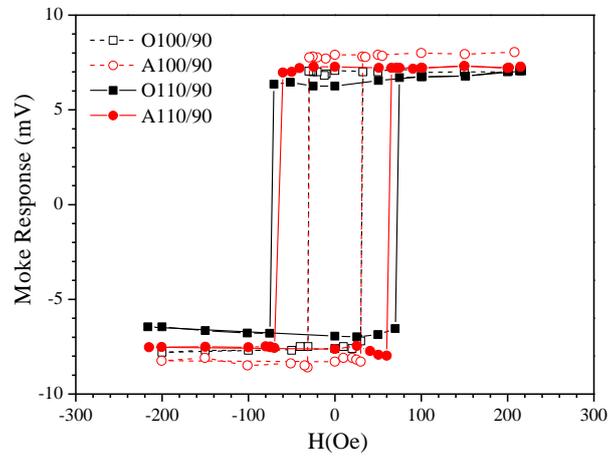

Figure 8